\documentclass{PoS}

\PoS{PoS(HEP2005)190}

\title{Running CP-violating Phases of Majorana Neutrinos}

\ShortTitle{Running CP-violating Phases of Majorana Neutrinos}

\author{\speaker{Zhi-zhong Xing}\thanks{In collaboration with S. Luo and
J.W. Mei (hep-ph/0507065), and supported in part by the National
Natural Science Foundation of China}\\
       Institute of High Energy Physics, Chinese Academy of Sciences,
       Beijing 100049, China\\
       E-mail: \email{xingzz@mail.ihep.ac.cn}}

\abstract{Three CP-violating phases of the $3\times 3$ lepton
flavor mixing matrix $V$, denoted as $\delta$, $\rho$ and $\sigma$
in a useful phase convention, are entangled with one another in
the one-loop renormalization-group equations. We show that $\delta
=90^\circ$ at the electroweak scale ($\Lambda_{\rm EW} \sim
10^{2}$ GeV) can be radiatively generated from $\delta =0^\circ$
at the seesaw scale ($\Lambda_{\rm SS} \sim 10^{14}$ GeV) in the
minimal supersymmetric standard model, provided three neutrino
masses are nearly degenerate. As for the Majorana phases $\rho$
and $\sigma$, it is also possible to radiatively generate $\rho
=90^\circ$ or $\sigma = 90^\circ$ at $\Lambda_{\rm EW}$ from $\rho
=0^\circ$ or $\sigma = 0^\circ$ at $\Lambda_{\rm SS}$. This
observation opens a new and interesting window for model building,
in order to understand possible connection between the phenomena
of CP violation at low- and high-energy scales.}

\FullConference{International Europhysics Conference on High Energy Physics\\
          July 21st - 27th 2005\\
          Lisboa, Portugal}

\begin{document}

Recent solar and atmospheric neutrino oscillation experiments have
provided us with very robust evidence that neutrinos are massive
and lepton flavors are mixed. A useful parametrization of the
$3\times 3$ lepton flavor mixing matrix $V$ reads \cite{FX01}:
\begin{equation}
V = \left( \matrix{ c^{}_{12}c^{}_{13} & s^{}_{12}c^{}_{13} &
s^{}_{13} \cr -c^{}_{12}s^{}_{23}s^{}_{13} - s^{}_{12}c^{}_{23}
e^{-i\delta} & -s^{}_{12}s^{}_{23}s^{}_{13} + c^{}_{12}c^{}_{23}
e^{-i\delta} & s^{}_{23}c^{}_{13} \cr -c^{}_{12}c^{}_{23}s^{}_{13}
+ s^{}_{12}s^{}_{23} e^{-i\delta} & -s^{}_{12}c^{}_{23}s^{}_{13} -
c^{}_{12}s^{}_{23} e^{-i\delta} & c^{}_{23}c^{}_{13} } \right)
\left ( \matrix{e^{i\rho } & 0 & 0 \cr 0 & e^{i\sigma} & 0 \cr 0 &
0 & 1 \cr} \right ) \; ,
\end{equation}
where $c^{}_{ij} \equiv \cos\theta_{ij}$ and $s^{}_{ij} \equiv
\sin\theta_{ij}$. Current data yield $\theta_{12} \approx
33^\circ$, $\theta_{23} \approx 45^\circ$ and $\theta_{13} <
10^\circ$ \cite{Fit}, but $\delta$, $\rho$ and $\sigma$ remain
entirely unrestricted. Many new neutrino experiments are underway,
not only to measure $\theta_{13}$ and $\delta$, but also to
constrain the Majorana phases $\rho$ and $\sigma$.

While neutrino masses and lepton flavor mixing parameters can be
measured at low-energy scales, their origin is most likely to
depend on some unspecified interactions at a superhigh energy
scale. For instance, the existence of very heavy right-handed
neutrinos and lepton number violation may naturally explain the
smallness of left-handed neutrino masses via the famous seesaw
mechanism at the scale $\Lambda_{\rm SS} \sim 10^{14}$ GeV. Below
this seesaw scale, the effective Lagrangian for lepton Yukawa
interactions can be written as $-{\cal L} = \overline{E_L^{}}H_1
Y_l^{} l_R^{} - \overline{E_L^{}} H_2 (\kappa/2) H_2^{c\dag} E_L^c
+ {\rm h.c.}$ in the minimal supersymmetric standard model (MSSM).
After spontaneous gauge symmetry breaking at the electroweak scale
$\Lambda_{\rm EW} \sim 10^2$ GeV, we arrive at the charged lepton
mass matrix $M_l = {\bf v}Y_l\cos\beta$ and the effective neutrino
mass matrix $M_\nu = {\bf v}^2\kappa \sin^2\beta$. The lepton
flavor mixing matrix $V$ arises from the mismatch between the
diagonalizations of $Y_l$ (or $M_l$) and $\kappa$ (or $M_\nu$). In
the flavor basis where $Y_l$ is diagonal, $V$ directly links the
neutrino mass and flavor eigenstates. The physical parameters at
$\Lambda_{\rm SS}$ and $\Lambda_{\rm EW}$ are related by the
renormalization group equations (RGEs). It has been shown that the
RGE evolution between these two scales may have significant
effects on the mixing angle $\theta_{12}$ and the phase parameters
$\delta$, $\rho$ and $\sigma$, and a CP-violating phase can even
be radiatively generated \cite{Casas}.

The reason for the radiative generation of a CP-violating phase is
simply that three phases of $V$ are entangled with one another in
the RGEs. Hence a non-zero value of $\delta$ can be generated at
$\Lambda_{\rm EW}$ from $\delta =0$ at $\Lambda_{\rm SS}$,
provided the initial values of $\rho$ and $\sigma$ are not
vanishing. This observation opens a new and interesting window to
understand possible connection between the phenomena of CP
violation at low- and high-energy scales; e.g., the phase
parameter governing the strength of CP violation in neutrino
oscillations could be radiatively generated from those
CP-violating phases which control the leptogenesis of right-handed
neutrinos at $\Lambda_{\rm SS}$. The purpose of this talk is just
to illustrate the {\it maximal} radiative generation of $\delta$,
$\rho$ and $\sigma$ in the RGE evolution from $\Lambda_{\rm SS}$
to $\Lambda_{\rm EW}$.

Below $\Lambda_{\rm SS}$, the one-loop RGE of $\kappa$ in the MSSM
can be found in Ref. \cite{Casas}. In the flavor basis where $Y_l$
is diagonal, we have $\kappa = V \overline{\kappa} V^T$ with
$\overline{\kappa} = {\rm Diag}\{\kappa^{}_1, \kappa^{}_2,
\kappa^{}_3\}$. The neutrino masses at $\Lambda_{\rm EW}$ read as
$m^{}_i = {\bf v}^2 \kappa^{}_i \sin^2\beta$ (for $i=1,2,3$). It
is then possible to derive the RGEs for $(\kappa^{}_1,
\kappa^{}_2, \kappa^{}_3)$, $(\theta_{12}, \theta_{23},
\theta_{13})$ and $(\delta, \rho, \sigma)$ \cite{Luo}. For
simplicity, here we only quote the approximate RGEs of $\delta$,
$\rho$ and $\sigma$ in the working assumption that three neutrino
masses are nearly degenerate \cite{Luo}:
\begin{eqnarray}
\frac{{\rm d} \delta}{{\rm d} t} & \approx &
\frac{y^2_\tau}{16\pi^2} \left [\frac{c^{}_{(\rho -\sigma)}
s^{}_{(\rho -\sigma)}}{\zeta^{}_{12}} s^2_{23} + \left (
\frac{c^{}_{(\delta -\rho)} s^{}_\rho}{\zeta^{}_{13}} -
\frac{c^{}_{(\delta -\sigma)} s^{}_\sigma}{\zeta^{}_{23}} \right )
\frac{c^{}_{12} s^{}_{12} c^{}_{23} s^{}_{23}}{s^{}_{13}} \right ]
\; ,
\nonumber \\
\frac{{\rm d} \rho}{{\rm d} t} & \approx &
\frac{y^2_\tau}{16\pi^2} \left [\frac{c^{}_{(\rho -\sigma)}
s^{}_{(\rho -\sigma)}}{\zeta^{}_{12}} s^2_{12} s^2_{23} + \left (
\frac{c^{}_{(\delta -\rho)} s^{}_\rho}{\zeta^{}_{13}} -
\frac{c^{}_{(\delta -\sigma)} s^{}_\sigma}{\zeta^{}_{23}} \right )
\frac{c^{}_{12} s^{}_{12} c^{}_{23} s^{}_{23}}{s^{}_{13}} \right ]
\; ,
\nonumber \\
\frac{{\rm d} \sigma}{{\rm d} t} & \approx &
\frac{y^2_\tau}{16\pi^2} \left [\frac{c^{}_{(\rho -\sigma)}
s^{}_{(\rho -\sigma)}}{\zeta^{}_{12}} c^2_{12} s^2_{23} + \left (
\frac{c^{}_{(\delta -\rho)} s^{}_\rho}{\zeta^{}_{13}} -
\frac{c^{}_{(\delta -\sigma)} s^{}_\sigma}{\zeta^{}_{23}} \right )
\frac{c^{}_{12} s^{}_{12} c^{}_{23} s^{}_{23}}{s^{}_{13}} \right ]
\; ,
\end{eqnarray}
where $\zeta^{}_{ij} \equiv (\kappa^{}_i -
\kappa^{}_j)/(\kappa^{}_i + \kappa^{}_j)$, $c^{}_{(\rho -\sigma)}
\equiv \cos (\rho -\sigma)$, $s^{}_{(\rho -\sigma)} \equiv \sin
(\rho -\sigma)$, etc. One can see that the one-loop RGE running
behaviors of $\delta$, $\rho$ and $\sigma$ are very similar to one
another.

I proceed to show a few typical numerical examples. The
eigenvalues of $Y_l$ at $\Lambda_{\rm SS}$ are chosen in such a
way that they can correctly run to their low-energy values. We
assume the masses of three light neutrinos to be nearly degenerate
and $m^{}_1 \sim 0.2$ eV, so as to make the RGE running effects of
relevant physical quantities significant enough. The initial
values of $\kappa^{}_i$ can be adjusted via $\kappa^{}_i =
m^{}_i/({\bf v}^2 \sin^2\beta)$ together with a typical input
$\tan\beta = 10$, such that the resultant neutrino mass-squared
differences $\Delta m^2_{21} \equiv m^2_2 - m^2_1$ and $\Delta
m^2_{31} \equiv m^2_3 - m^2_1$ at $\Lambda_{\rm EW}$ are
consistent with the solar and atmospheric neutrino oscillation
data. We follow a similar strategy to choose the initial values of
three mixing angles $\theta_{12}$, $\theta_{23}$ and
$\theta_{13}$, in order to reproduce their low-energy values
determined or constrained by current experimental data \cite{Fit}.
In view of the upper bound $\theta_{13} < 10^\circ$, we shall
typically take $\theta_{13} = 1^\circ$, $3^\circ$ and $5^\circ$ in
our numerical calculations. We allow one of three CP-violating
phases to vanish at $\Lambda_{\rm SS}$ and examine whether it can
run to $90^\circ$ at $\Lambda_{\rm EW}$ by choosing the initial
values of the other two phase parameters properly.

{\it (A) Radiative generation of $\delta=90^\circ$.} ~ The input
and output values of relevant parameters are listed in Table 1. We
see that $\delta = 90^\circ$ at $\Lambda_{\rm EW}$ can be
radiatively generated from $\delta = 0^\circ$ at $\Lambda_{\rm
SS}$, if $\theta_{13} = 1^\circ$, $\rho = 4.0^\circ$ and $\sigma =
-57.5^\circ$ are input. Changing the initial value of
$\theta_{13}$ to $3^\circ$ or $5^\circ$ but fixing the input
values of the other quantities, we find that only $\delta =
41.8^\circ$ or $\delta = 35.8^\circ$ can be obtained at
$\Lambda_{\rm EW}$. While the results of $m^{}_1$, $\Delta
m^2_{31}$, $\theta_{12}$ and $\theta_{23}$ at $\Lambda_{\rm EW}$
are rather stable against the change of $\theta_{13}$ from
$1^\circ$ to $5^\circ$ at $\Lambda_{\rm SS}$, the result of
$\Delta m^2_{21}$ becomes smaller and less favored.
\begin{table}
\caption{Radiative generation of $\delta = 90^\circ$ in the MSSM
with $\tan\beta = 10$.} \vspace{0.2cm}
\begin{center}
\begin{tabular}{c|l|ccc} \hline\hline
Parameter & Input $\left(\Lambda_{\rm SS}^{}\right)$ ~~~~~~ &
\multicolumn{3}{c}{Output $( \Lambda_{\rm EW} )$} \\ & &
$\theta_{13}
= 1^\circ$ & $\theta_{13} = 3^\circ$ & $\theta_{13} = 5^\circ$ \\
\hline
$m^{}_1 ({\rm eV} )$ & 0.241 & 0.20 & 0.20 & 0.20 \\
$\Delta m^2_{21} ( 10^{-5} ~{\rm eV}^2 )$ & 20.4 & 7.79 & 7.17 & 6.56 \\
$\Delta m^2_{31} ( 10^{-3} ~{\rm eV}^2 )$ & 3.32 & 2.20 & 2.20 & 2.20 \\
\hline
$\theta_{12}$ & $24.1^\circ$ & $33.0^\circ$ & $33.0^\circ$ & $33.1^\circ$ \\
$\theta_{23}$ & $43.9^\circ$ & $45.1^\circ$ & $45.0^\circ$ & $45.0^\circ$ \\
$\theta_{13}$ & $1^\circ/3^\circ/5^\circ$ & $0.65^\circ$ & $2.46^\circ$
& $4.52^\circ$ \\
\hline
$\delta$ & $0^\circ$ & $90.0^\circ$ & $41.8^\circ$ & $35.8^\circ$ \\
$\rho$ & $4.0^\circ$ & $72.2^\circ$ & $23.8^\circ$ & $17.6^\circ$ \\
$\sigma$ & $-57.5^\circ$ & $26.3^\circ$ & $-22.0^\circ$ &
$-28.1^\circ$ \\ \hline\hline
\end{tabular}
\end{center}
\end{table}

Next let us examine whether the radiative generation of $\delta =
90^\circ$ at $\Lambda_{\rm EW}$ can be achieved from other initial
values of $\rho$ and $\sigma$, when $\theta_{13} = 1^\circ$ holds
and $m^{}_1$, $\Delta m^2_{21}$, $\Delta m^2_{31}$, $\theta_{12}$
and $\theta_{23}$ take the same input values as before at
$\Lambda_{\rm SS}$ (see Table 1). We find out two new numerical
examples with $(\rho, ~\sigma) = (0^\circ, -59.9^\circ)$ and
$(10^\circ, -57^\circ)$, respectively. Note that both $\delta$ and
$\rho$ (or $\sigma$) can be radiatively generated from $\sigma
\neq 0^\circ$ (or $\rho \neq 0^\circ$) at the seesaw scale. The
possibility to simultaneously generate $\rho$ and $\sigma$ from
$\delta \neq 0^\circ$ at $\Lambda_{\rm SS}$ via the RGE evolution
is in general expected to be strongly suppressed, because the
leading terms of ${\rm d}\rho/{\rm d}t$ and ${\rm d}\sigma/{\rm
d}t$ in Eq. (2) vanish for $\rho = \sigma = 0^\circ$ at
$\Lambda_{\rm SS}$. Only in the $\theta_{13} \rightarrow 0$ limit,
the running effects of three CP-violating phases could become
significant. Our results hint at the existence of strong parameter
degeneracy in obtaining $\delta = 90^\circ$ at $\Lambda_{\rm EW}$
from $\delta = 0^\circ$ at $\Lambda_{\rm SS}$. To resolve this
problem is certainly a big challenge in model building, unless two
Majorana CP-violating phases could separately be measured at low
energies.

{\it (B) Radiative generation of $\rho=90^\circ$ or
$\sigma=90^\circ$.} ~ Now we take a look at the one-loop RGE
evolution of $\rho$. The inputs and outputs of relevant parameters
are listed in Table 2. One can see that there is no difficulty to
radiatively generate $\rho = 90^\circ$ at $\Lambda_{\rm EW}$ from
$\rho = 0^\circ$ at $\Lambda_{\rm SS}$, provided $\theta_{13} =
1^\circ$, $\delta = 0^\circ$ and $\sigma = -67.7^\circ$ are input.
Allowing the initial value of $\theta_{13}$ to change to $3^\circ$
or $5^\circ$ but fixing the input values of the other quantities,
we find that only $\rho = 17.6^\circ$ or $\rho = 12.1^\circ$ can
be obtained at $\Lambda_{\rm EW}$. Again the results of $m^{}_1$,
$\Delta m^2_{31}$, $\theta_{12}$ and $\theta_{23}$ at
$\Lambda_{\rm EW}$ are quite stable against the change of
$\theta_{13}$ from $1^\circ$ to $5^\circ$ at $\Lambda_{\rm SS}$,
but the result of $\Delta m^2_{21}$ becomes smaller.
\begin{table}
\caption{Radiative generation of $\rho = 90^\circ$ in the MSSM
with $\tan\beta = 10$.} \vspace{0.2cm}
\begin{center}
\begin{tabular}{c|l|ccc} \hline\hline
Parameter & Input $\left(\Lambda_{\rm SS}^{}\right)$ ~~~~~ &
\multicolumn{3}{c}{Output $( \Lambda_{\rm EW}
)$} \\
& & $\theta_{13}
= 1^\circ$ & $\theta_{13} = 3^\circ$ & $\theta_{13} = 5^\circ$ \\
\hline
$m^{}_1 ({\rm eV} )$ & 0.241 & 0.20 & 0.20 & 0.20 \\
$\Delta m^2_{21} ( 10^{-5} ~{\rm eV}^2 )$ & 20.4 & 8.54 & 7.90 & 7.27 \\
$\Delta m^2_{31} ( 10^{-3} ~{\rm eV}^2 )$ & 3.32 & 2.21 & 2.20 & 2.20 \\
\hline
$\theta_{12}$ & $27.6^\circ$ & $33.1^\circ$ & $33.2^\circ$ & $33.3^\circ$ \\
$\theta_{23}$ & $43.9^\circ$ & $44.8^\circ$ & $44.8^\circ$ & $44.8^\circ$ \\
$\theta_{13}$ & $1^\circ/3^\circ/5^\circ$ & $0.43^\circ$ & $2.17^\circ$
& $4.24^\circ$ \\
\hline
$\delta$ & $0^\circ$ & $107.6^\circ$ & $35.4^\circ$ & $30.2^\circ$ \\
$\rho$ & $0^\circ$ & $90.2^\circ$ & $17.6^\circ$ & $12.1^\circ$ \\
$\sigma$ & $-67.7^\circ$ & $34.1^\circ$ & $-38.3^\circ$ &
$-44.6^\circ$ \\ \hline\hline
\end{tabular}
\end{center}
\end{table}

In a similar way, one may realize the radiative generation of
$\sigma=90^\circ$.

\vspace{0.2cm}

Finally it is worth remarking that our analysis is essentially
independent of the specific textures of lepton Yukawa coupling
matrices. Thus it can be applied to the concrete work of model
building. Since the elegant thermal leptogenesis mechanism is
usually expected to work at the seesaw scale, a study of its
consequences at low-energy scales is available by means of the
one-loop RGEs that we have obtained. In other words, the RGEs may
serve as a useful bridge to establish a kind of connection between
the phenomena of CP violation at low- and high-energy scales.

\vspace{0.2cm}

I apologize for not being able to cite more relevant works, just
due to the page limitation.


\begin{thebibliography}{99}
\bibitem{FX01} H. Fritzsch and Z.Z. Xing, Phys. Lett. B {\bf 517},
363 (2001); Z.Z. Xing, Int. J. Mod. Phys. A {\bf 19}, 1 (2004).

\bibitem{Fit} A. Strumia and F. Vissani, hep-ph/0503246.

\bibitem{Casas} J.A. Casas {\it et al.}, Nucl. Phys. B {\bf 573}, 652 (2000).
S. Antusch {\it et al.}, Nucl. Phys. B {\bf 674}, 401 (2003).

\bibitem{Luo} S. Luo, J.W. Mei, and Z.Z. Xing, hep-ph/0507065; Phys. Rev. D
(in press).
\end{thebibliography}
\end{document}